\documentclass[a4paper,11pt]{article}
\pdfoutput=1 % if your are submitting a pdflatex (i.e. if you have
\usepackage{jcappub} % for details on the use of the package, please
\usepackage[T1]{fontenc} % if needed
\graphicspath{{images/}}
\usepackage{diagbox}
\usepackage{float}
\usepackage{color}

\title{\boldmath On the contribution of cosmic-ray interactions in the circumgalactic gas to the observed high-energy neutrino flux}

\author[a,b]{Oleg Kalashev,}
\author[c,a,1]{Nickolay Martynenko,\note{Corresponding author.}}
\author[a,c]{and Sergey Troitsky}

\affiliation[a]{Institute for Nuclear Research of the Russian Academy of Sciences,\\ 60th October Anniversary Prospect 7a, Moscow 117312, Russia}
\affiliation[b]{Moscow Institute for Physics and Technology,\\ 9 Institutskiy per., Dolgoprudny, Moscow Region, 141701 Russia}
\affiliation[c]{Faculty of Physics, M.V. Lomonosov Moscow State University,\\ 1-2 Leninskie Gory, Moscow 119991, Russia}

\emailAdd{kalashev@ms2.inr.ac.ru}
\emailAdd{martynenko.ns18@physics.msu.ru}
\emailAdd{st@ms2.inr.ac.ru}

\abstract{Cosmic rays escaping the Milky-Way disk interact with circumgalactic gas which fills the virial volume of our Galaxy. These interactions should produce guaranteed fluxes of energetic diffuse neutrinos and photons observable at the Earth. This neutrino flux would be a plausible contribution to the spectrum measured by the IceCube neutrino observatory: the energy emitted in this way is weakly constrained from cascade gamma rays, since the cascades have no time to develop, but the arrival directions of the neutrinos do not point to the Galactic disk, in agreement with observations. However, previous studies reported very different estimates of the corresponding neutrino flux, so it was unclear if this contribution to the observed spectrum is essential. Here we readdress the calculation of this diffuse neutrino flux component under various assumptions about the cosmic-ray spectrum and propagation in the circumgalactic medium. We find that even with these variations, this contribution to the observed neutrino flux remains subleading provided multimessenger constraints are satisfied.}


\arxivnumber{2207.12458}

\begin{document}
\maketitle
\flushbottom

\section{Introduction}
\label{sec:intro}
First reported nine years ago \cite{IceCube-2013}, the presence of extraterrestrial high-energy neutrinos is now confirmed at the confidence level above 5$\sigma$ by IceCube \cite{IceCube-HESE-2020}, $\sim 2 \sigma$ by ANTARES \cite{ANTARES-diffuse} and $3\sigma$ by Baikal-GVD \cite{Baikal-Neutrino2022a,Baikal:2022chp}. However, the origin of these astrophysical neutrinos remains uncertain, and simple models do not succeed in explaining all observations (for a recent review, see e.g.\ ref.~\cite{ST-UFN}). In particular, the dominant part of astrophysical neutrinos do not point back to the Galactic plane \cite{IceCubeANTARES-GalPlane}, which suggests their extragalactic origin. At the same time, the spectrum of astrophysical neutrinos below $\sim 100$~TeV, measured by IceCube in the cascade mode, is in a tension with the hypothesis of 100\% extragalactic origin, considered in the multimessenger framework, see ref.~\cite{IceCube-HESE-2020} and references therein. This is because the energetic gamma rays, accompanying neutrinos in all standard production mechanisms, experience electromagnetic cascades on cosmic background photons \cite{Nikishov1962,BerezinskyKalashev}, and secondary lower-energy gamma rays may overshoot \cite{BerezinskySmirnov1975} the GeV-band isotropic gamma-ray flux measured by Fermi-LAT \cite{FermiDiffuse}. To overcome this tension, one may suppose that the sources are opaque to gamma-ray photons, but a large optical depth is then required because the energy of photons exiting the source should degrade down to the poorly explored MeV energy band, see e.g.\ ref.~\cite{MuraseCorona1}.

Another possible resolution to this tension might be provided by the Galactic origin of a part of neutrinos: in this case, their sources are so close that electromagnetic cascades do not have enough time to develop, and gamma rays reaching the Earth remain in the energy band above the Fermi-LAT sensitivity, see e.g.\ refs.~\cite{AhlersMurase-gamma-Gal,KalashevST-gamma-Gal}. A combination of Galactic and extragalactic contributions \cite{2comp-Chen,2comp-Vissani,2comp-Neronov,2comp-Vissani-2} may help to explain the apparent tension between neutrino spectra measured by IceCube in the cascade and track channels, see ref.~\cite{ST-UFN} for a recent discussion. This is backed up by the fact that the contribution of radio blazars, which correlate with IceCube \cite{neutradio1,hovatta,neutradio2} and ANTARES \cite{ANTARES-radioJulien, ANTARES-radioGiulia} events, may explain the entire muon-track IceCube flux \cite{neutradio2}, leaving some space for an additional lower-energy contribution possibly seen in the cascade spectrum. Interestingly, observational indications to the presence of the Galactic component were recently found in the distribution of arrival directions of neutrino events detected by IceCube \cite{neutGalaxy} and ANTARES \cite{ANTARES-GalRidge}. 

The Galactic contribution, if essential, should be consistent with the overall isotropy in the neutrino arrival directions. This may happen in two cases: either the dominant part of the flux comes from immediate neighbourhood of the Solar system, that is from the region smaller than the Galactic-disk thickness \cite{NeronovBubble2018,LocalBubble2020}, or it is collected from a spherical halo of the size much larger than the disk \cite{Taylor:halo-neutrino}. Of particular interest is the scenario when the neutrinos are produced in the huge halo of circumgalactic gas, which extends all the way to the virial radius of the Galaxy, that is about ten times the radius of the Milky-Way disk \cite{Gupta:2012-gas}. Cosmic protons are accelerated in the disk and, at sufficiently high energies, escape it and interact with the gas around the Galaxy. Energetic neutrinos and photons produced in these interactions contribute to the diffuse, full-sky backgrounds observed at the Earth. The importance of the gamma-ray contribution was stressed in ref.~\cite{Feldmann:halo-photons}, while the possible corresponding contribution to the neutrino background was mentioned in ref.~\cite{Taylor:halo-neutrino} without a detailed calculation. However, a subsequent numerical study \cite{KT2016}, assuming the diffusive escape of cosmic rays from the disk, demonstrated that only a tiny fraction of the IceCube neutrino flux can be explained in this way, provided the spectrum of cosmic rays in the disk does not exceed that observed at the Earth. Recently, ref.~\cite{Gabici:2021-crhalo} relaxed the assumption of the diffusive escape and assumed a very hard spectrum of cosmic rays in the Milky-Way halo, relating it to some interpretations of gamma-ray observations of another giant spiral, the Andromeda galaxy M31.
In the present work, we readdress in more detail the production of high-energy neutrinos in the circumgalactic gas for various assumptions about the cosmic-ray escape and propagation. The aim of our study is to explore variations of the diffusive-escape model of ref.~\cite{KT2016} and of the M31-inspired model of ref.~\cite{Gabici:2021-crhalo}, taking into account multimessenger constraints from local observations at the Earth.

The rest of the paper is organized as follows. In section~\ref{sec:anal}, we describe our calculations, presenting first the general approaches in section~\ref{sec:anal:methods} and section~\ref{sec:anal:cgm}, and then their applications to the diffusive (section~\ref{sec:anal:diff}) and non-diffusive (section~\ref{sec:anal:ah}) cosmic-ray escape scenarios. The results are presented and discussed in section~\ref{sec:results}. We briefly conclude in section~\ref{sec:concl}.

\section{Analysis}
\label{sec:anal}

\subsection{Methods}
\label{sec:anal:methods}
Following ref.~\cite{KT2016}, we calculate fluxes of secondary neutrinos and gamma rays from interactions of cosmic rays with circumgalactic gas by integrating contributions from all directions taking into account non-central position of the Sun in the Galaxy. Below we assume, for simplicity, that both cosmic rays and circumgalactic gas are dominated by protons, and neglect the contribution of heavier nuclei to the high-energy neutrino production. The source density for neutrino ($\nu$), photon ($\gamma$) and electron/positron ($e$) production,
\begin{equation}
 Q_{\mathrm{\nu, \gamma, e}}(E, r) = \int dE' \, n_{\rm{CGM}}(r) \frac{dn_{\rm{CR}}}{d E'}(E', r)\frac{d\sigma_{\rm{pp}}}{dE}(E'),
\end{equation}
is defined by the distribution of the circumgalactic gas, $n_{\rm{CGM}}(r)$, as well as the density and the spectrum of cosmic rays, $d n_{\rm{CR}}/d E'$. Here $d\sigma_{\rm{pp}}(E')/dE$ is the differential cross section of proton-proton interactions, $E'$ denotes the cosmic-ray proton energy and $E$ denotes the energy of the secondary photon or neutrino. For neutrinos, the flux calculation from a given direction reduces to integration of the source term over the line of sight,
\begin{equation}
    j_{\mathrm{\nu}}(E) = \int Q_{\mathrm{\nu}}(E,r(s)) ds.
\end{equation}
Since we aim to estimate the total flux of all neutrino flavors, we do not need to account for neutrino oscillations.

For gamma rays we also take into account the suppression of the flux due to the pair production process on cosmic microwave background (CMB) photons (with the secondary photons produced in the cascades taken into account) and extra contribution of secondary electrons via inverse Compton scattering on CMB.
This is done in a simplified way by solving one-dimensional transport equations with the numerical code \cite{Kalashev:2014xna}. We note that the attenuation on infrared/optical (IR/O) background is negligible within the Milky-Way halo, since the IR/O contribution to the pair production mean free path is significant only at the distances of $>1$~Mpc (see e.g.~refs.~\cite{2010ApJ...710.1530V,2016PhRvD..94b3007B}), which is much larger than the Galactic virial radius $R_{\rm{vir}}$. 
Then we average the secondary flux of photons produced via inverse Compton scattering to account for the isotropization of electrons. Finally, since we are interested in the isotropic part of the flux, we take the minimal total flux among all the directions, which is the flux from the Galactic anticenter, and treat it as the isotropic component. We normalize the secondary fluxes in such a way that the local gamma-ray flux does not exceed that observed by Fermi-LAT~\cite{FermiDiffuse} at all energies. We also consider a normalization to the local integrated gamma-ray flux observed by Fermi-LAT with point sources' contribution subtracted~\cite{Fermi-LAT:2015otn}. 

The constraint obtained in this way can be considered as conservative, since we do not take into account the extra contribution of secondary photons produced by electrons due to their increased travel path in the halo. We expect the correction to be less than factor of $\simeq 2$ since roughly equal energy is emitted in the form of electrons and photons.

\subsection{Circumgalactic gas}
\label{sec:anal:cgm}
We assume a commonly used, see e.g. ref.~\cite{2018ApJ...862....3B}, spherical $\beta$-model for the Milky-Way circumgalactic gas number density distribution,
\begin{equation}
    n_{\rm{CGM}}(r) = n_0 \left(1 + r^2/r_c^2\right)^{-3\beta/2},  
\end{equation}
where $r$ is the Galactocentric radius, $n_0$, $r_c$ and $\beta$ are the normalization, core radius and slope parameter, respectively. This parametrization was originally motivated by the fact that it reproduces the observed X-ray surface brightness profile of external galaxies (for more details, see e.g. ref.~\cite{Martynenko:2021} and references therein).
The value of $r_c\sim 1$~kpc is poorly constrained by observations, but this parameter almost does not affect the gas density at the region of interest ($r\gg1$~kpc). Speculatively, we hereafter fix $r_c=3$~kpc for our calculations.

At large galactocentric radii ($r\geq30$ kpc), we use the profile from ref.~\cite{Martynenko:2021}, where an isothermal hot ($T\simeq 2\times10^6$~K) gas near hydrostatic equilibrium was assumed, and a joint analysis of OVII X-ray spectra (both absorption and emission) and ram-pressure stripping of Milky-Way dwarf satellites data was provided with the gas metallicity gradient taken into account. The latter is significant for the spectral data analysis, since oxygen ions are only tracers of much more abundant gas, and thus it is important to estimate their fraction in the gas composition carefully. From the other hand, the use of the ram-pressure stripping data allows to probe the gas density independently from its chemical composition, although this sample is considerably smaller and less precise compared to the spectral one.

\begin{figure}[tpb]
    \centering
    \includegraphics[width=\columnwidth]{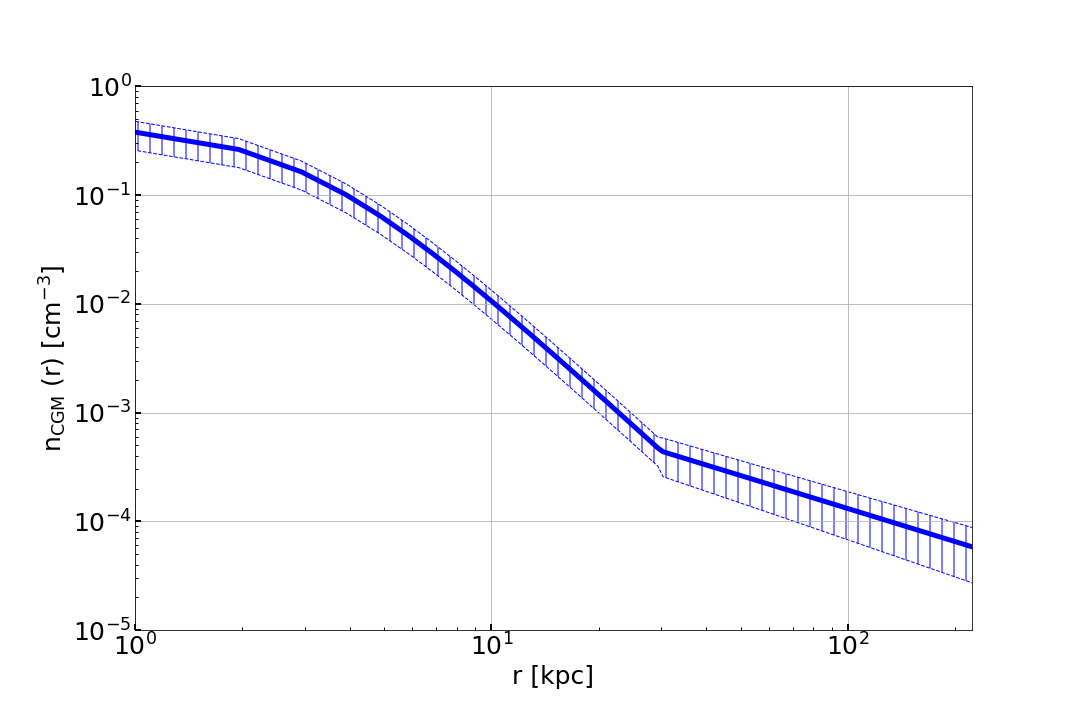}
    \caption{The circumgalactic gas density profile ($r<30$~kpc: ref.~\cite{Feldmann:halo-photons}, $r\geq30$~kpc: ref.~\cite{Martynenko:2021}). The hatched region corresponds to the 68\% confidence level uncertainty.
    }
    \label{fig:CGM-density}
\end{figure}

The discussed profile is however inapplicable at smaller distances, specifically, near the Milky-Way disk (for a detailed discussion, see ref.~\cite{Martynenko:2021}). Therefore, within $r<30$ kpc we use the result of hydrodynamical cosmological simulations describing the circumgalactic gas spatial and chemical evolution provided in ref.~\cite{Feldmann:halo-photons}.

 Also, it should be noted that the analysis in ref.~\cite{Martynenko:2021} probed only the spherical component of the gas density distribution (because of the data filtering procedure discussed therein). Therefore, the present study does not estimate the non-spherical component in neutrino and gamma-ray signals which is due to their production in the Galactic halo substructures (high-latitude clouds or other non-spherical features in the gas density distribution, see e.g. refs.~\cite{2016A&A...592A.142R,Haffner_2003}, respectively).

Thus we obtain for the normalization and slope parameter (we hereafter use the notation $x_{\text{unit}}\equiv x/\text{unit}$, e.g. $r_{\text{kpc}}\equiv r/\text{kpc}$; the uncertainties correspond to the $68\%$ confidence level interval):
\begin{subequations}
\begin{align}
        n_{0_{\text{cm}^{-3}}} = 4.54^{+1.17}_{-1.45} \times 10^{-3},~& \beta = 0.337^{+0.043}_{-0.028}, & r_{\text{kpc}}\geq 30\\
        n_{0_{\text{cm}^{-3}}} = 4.47^{+1.15}_{-1.43} \times 10^{-1},~& \beta = 1.000, & r_{\text{kpc}}<30
\end{align}
\end{subequations}
Since the parameters in the inner region were obtained from computer simulations and not from fitting the observations, we assumed here that the relative uncertainty in the normalization is equal to that for the outer region and did not estimate the slope uncertainty.

Figure~\ref{fig:CGM-density} presents the resulting circumgalactic gas density profile.
\subsection{Diffusive escape}
\label{sec:anal:diff} %(NM)
To date, there are a number of studies which suggest cosmic rays to diffusively escape from the Milky-Way disk to the circumgalactic medium~\cite{Feldmann:halo-photons,Taylor:halo-neutrino,KT2016}. The most recent result~\cite{KT2016} indicates that the circumgalactic contribution to the observed astrophysical neutrino flux does not exceed $\sim$~1\%. Nevertheless, this conclusion depends significantly on the assumed circumgalactic gas profile shape. 

In this work, the scenario with the diffusive cosmic-ray propagation is considered within the same approach that was detailed in ref.~\cite{KT2016}, but using the results from ref.~\cite{Martynenko:2021} to describe the gas profile shape at large galactocentric distances ($r>30$~kpc). In order to compare between diffusive and non-diffusive scenario correctly, we also use a hard injection spectrum of $E'^{-2}$ and localize the source within $r<15$ kpc. %The latter allows us to account for any type of cosmic-ray-producing activity near the Galactic disk and Fermi bubbles.

The proton spectral density in the Milky-Way circumgalactic medium $j(E', r, t) \equiv dn_{\mathrm{CR}}/dE'$ is obtained by solving the diffusion equation:
\begin{equation}
        \left[\partial_t - D(E') \Delta_r + c\sigma_{\mathrm{pp}} (E') n_{\mathrm{CGM}}(r)\right]j (E', r, t) = Q_{\mathrm{p}}(E', r, t),
    \label{eq:diffusion}
\end{equation}
where $E'$, $r$, and $t$ are the cosmic-ray proton energy, Galactocentric radius, and time, respectively, $D(E') = D_{0} (E'/\mathrm{GeV})^{1/3}$ is the diffusion coefficient (the Kolmogorov turbulence regime is assumed), $\Delta_r \equiv r^{-2} \partial_r (r^2 \partial_r)$ is the radial part of the three-dimensional Laplace operator, $c$ is the speed of light, $\sigma_{\mathrm{pp}}(E')$ is the proton-proton interaction cross-section, $n_{\mathrm{CGM}}(r)$ is the proton number density in the circumgalactic medium, and $Q_{\mathrm{p}}(E', r, t)$ is the source term (the spectral density of the cosmic-ray density injection rate).

In order to obtain $\sigma_{\mathrm{pp}}(E')$, we parametrize the function as $\sigma_{\mathrm{pp}}(E') = \sigma_0 + \sigma_1 (\log E'/{\text{GeV}})$ $+\sigma_2 (\log E'/{\text{GeV}})^2$ and fit the coefficients $\sigma_{i}$ to the ref.~\cite{PDG:2022} data in the energy range of $E'>10$~GeV.

Following~ref.~\cite{KT2016}, we adopt the value of $D_0 = 1.2\times10^{29}$ cm$^{2}$ s$^{-1}$ originally used in ref.~\cite{Feldmann:halo-photons} for the diffusion coefficient. It is notable that the actual value of the coefficient depends on poorly constrained magnetic field in the circumgalactic medium. However, in ref.~\cite{KT2016}, it has conclusively been shown that the variations in $D_0$ by an order of magnitude do not change the resulting neutrino flux qualitatively. Thus, we hereafter fix the adopted $D_0$ value and do not explore how its variations affect our results.

For the source term, we adopt the same shape as in ref.~\cite{KT2016}, but with a harder spectrum and a larger localization radius:
\begin{equation}
    Q_{\mathrm{p}}(E', r, t) \propto  E'^{-\alpha} \exp\left(\frac{-E'}{E'_{\mathrm{cut}}}\right)\theta(r_{\mathrm{Q}} - r) \times Q_{\mathrm{time}}(t),
    \label{eq:sourceterm}
\end{equation}
where $\alpha = 2$, $E'_{\mathrm{cut}} = 10^8$ GeV are the injection spectrum parameters, $\theta$ is the step function, $r_{\mathrm{Q}} = 15$ kpc is the source localization radius, and ``simple'' function $Q_{\mathrm{time}}(t)$ describes the evolution of the source (in exactly the same way as in refs.~\cite{Feldmann:halo-photons,KT2016}; $t$ denotes the time from the final assembly of the inner Galactic halo):
\begin{equation}
    Q_{\mathrm{time}}(t) = \begin{cases}1 + t_{\text{Gyr}},~t_{\text{Gyr}} \in[0,2) \\
        3,~ t_{\text{Gyr}}\in[2,6) \\
        6 - t_{\text{Gyr}}/2,~t_{\text{Gyr}}\in[6,10]
    \end{cases}
    \label{eq:sourceterm_evolution}
\end{equation}
Let us note that the spatial structure of the source responsible for delivering the cosmic rays to the Galactic halo is not yet understood, and thus the source is assumed to be uniformly distributed within $r<15$~kpc. Although this volume is only $\sim 10^{-4}$ of the total Galactic halo one, the assumption allows us to account for any type of cosmic-ray-producing activity near the Galactic disk and Fermi bubbles. Moreover, we expect the variations in the source structure within $r<15$~kpc to have insignificant impact on the results (e.g., we obtain a negligible change in our results assuming a constant source density per unit phase volume, $\theta(r_{Q}-r)/r^2$, or considering a smaller localization radius of $r_{\mathrm{Q}} = 5$~kpc or $r_{\mathrm{Q}} = 10$~kpc). Thus, our assumption is reasonable to be adopted.

\begin{figure}[tpb]
    \centering
    \includegraphics[width=\columnwidth]{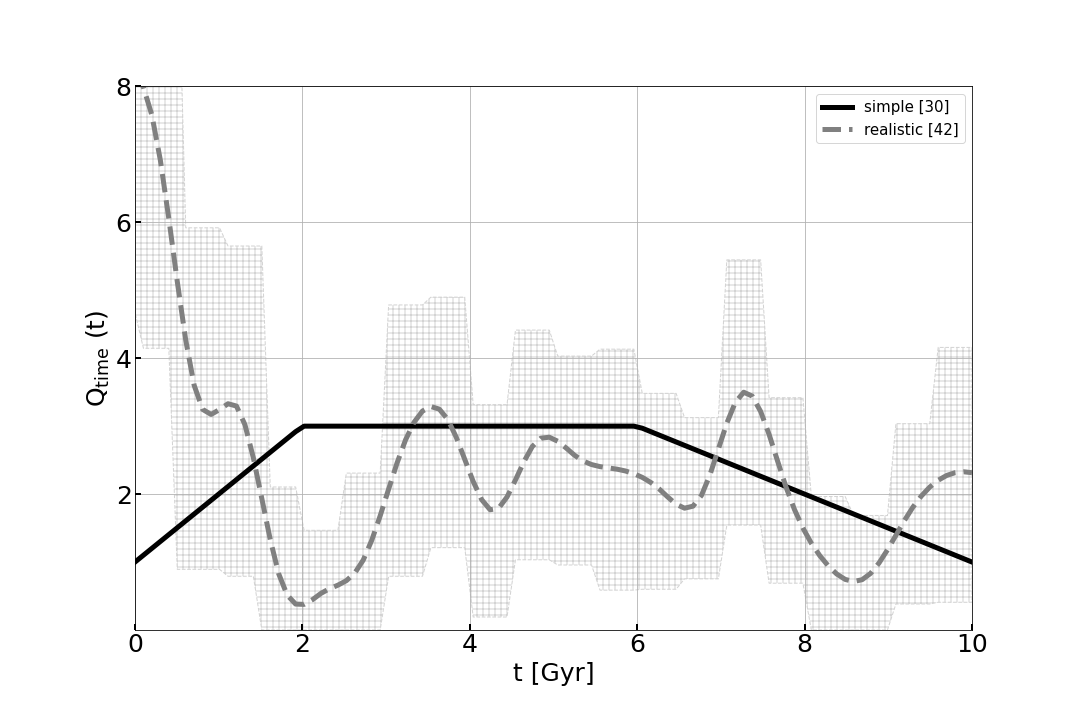}
    \caption{The source temporal evolution for the ``simple'' and ``realistic'' model (refs.~\cite{Feldmann:halo-photons,2015A&A...578A..87S}, respectively). For the ``realistic'' model, 68\% confidence level region is shown.}
    \label{fig:Qtime}
\end{figure}

The assumed temporal dependence~\eqref{eq:sourceterm_evolution} is originally intended to reflect variations in the star formation rate (SFR) of the Milky-Way~\cite{Feldmann:halo-photons}. Here we adopt the same dependence as in previous studies so that all inconsistencies between our and their results would be from taking into account the updated knowledge on the gas distribution and assuming a harder spectrum. It should be however taken in mind that the temporal evolution of the cosmic-ray source function is poorly constrained, and this implies a systematic uncertainty in the resulting fluxes, which is not estimated here. In order to test whether the assumed temporal dependence is significant for our results, we also consider another, ``realistic'' function $Q_{\mathrm{time}}(t)$ which reflects the best-fit Milky-Way SFR evolution reproducing silicon to iron abundance ratio in the Galactic disk from ref.~\cite{2015A&A...578A..87S}. We normalize the ``realistic'' function to have the same value of $\int dt~ Q_{\mathrm{time}}(t)$ as that of the ``simple'' function defined in~\eqref{eq:sourceterm_evolution}. 
Figure~\ref{fig:Qtime} compares the two functions. Let us note that the functions are consistent within the uncertainty at $t_{\text{Gyr}}>3$. 
We find the difference between these two source evolution scenarios in terms of the present-day local gamma-ray and neutrino fluxes to be negligible, and thus only the ``simple'' model results are hereafter presented.

We solve eq.~\eqref{eq:diffusion} numerically using \texttt{numpy}, \texttt{scipy} packages~\cite{Numpy:2020,Scipy:2020} (see appendix~\ref{app:numsol} for a detailed discussion of the numerical solution). The obtained present-day ($t_{\text{Gyr}}=10$) proton spectral density $j(E', r)$ is then used to calculate the local photon and 6-neutrino fluxes in the way described in section~\ref{sec:anal:methods}. 

\subsection{Non-diffusive escape} %(NM)
\label{sec:anal:ah}

In ref.~\cite{Gabici:2021-crhalo}, the authors argue the diffusive escape scenario to be inconsistent with the M31 observations presented in ref.~\cite{Karwin:2019}. These observations include gamma-ray intensity measurements for the outer and intermediate regions of the M31 halo. In the analysis from ref.~\cite{Gabici:2021-crhalo}, it is concluded that the ratio between these intensities implied by the diffusive scenario differs from that observed by, at least, an order of magnitude.
Nevertheless, the alternative scenarios proposed in ref.~\cite{Gabici:2021-crhalo} which can complement or substitute the diffusion are not studied thoroughly enough to derive the cosmic-ray density profile and spectrum and (in particular, due to the lack of observations) to single out one of the mechanisms.

Instead of considering a specific alternative escape scenario (see e.g. ref.~\cite{Roy:2022}), in the current analysis, we estimate the non-diffusive cosmic-ray density profile directly from the M31 observations presented in ref.~\cite{Karwin:2019} and used in the argumentation of ref.~\cite{Gabici:2021-crhalo}. 

It should be noted that the term ``non-diffusive'' does not imply negligible diffusion, but points to an essential contribution of other processes to the cosmic-ray transport. In addition, it should be noted that our approach to constrain the cosmic-ray density profile is qualitative and the results should be interpreted with caution. However, this approach allows us to obtain useful results with a limited number of simple assumptions consistent with observations.

The gamma-ray emission of the ``Northern'' (in terms of the Galactic latitude) hemisphere of the M31 halo is significantly contaminated by the Milky-Way disk contribution, so we use only the data from the ``Southern'' hemisphere, assuming the M31 halo to be symmetric (see Table~13 in ref.~\cite{Karwin:2019}).
For the further analysis, we use the observed ratio of the ``Spherical Halo'' and ``Far Outer Halo'' gamma-ray intensities. The former region corresponds to $r\simeq (5...116)$~kpc, and the latter to $r\simeq (116...219)$~kpc. The observed ratio is $I_{\mathrm{SH}} / I_{\mathrm{FOH}} \simeq 1.4\pm1.0$ (note that this value corresponds to the photon energies of $E<100$~GeV, while the further analysis focuses on considerably higher energies, $E\sim100$~TeV).

For simplicity, let us assume that the cosmic-ray proton spectral density $j(E', r)$ factorizes as $n_{\mathrm{CR}}(r) \times \left(E'/\text{GeV}\right)^{-\alpha} \exp\left(-E'/E'_{\mathrm{cut}}\right)$. To be consistent with ref.~\cite{Gabici:2021-crhalo}, for the spectral part, we hereafter adopt $\alpha=2$ and $E'_{\mathrm{cut}}= 2\times10^7$~GeV. 

Under the assumptions discussed above, see section~\ref{sec:anal:methods}, the gamma-ray source density is proportional to the product of number densities of the halo gas and of cosmic-ray protons. ref.~\cite{Gabici:2021-crhalo} did not study the radial dependence of this source density, effectively assuming it is constant throughout the halo. To work with a more realistic, though still a toy-model profile, we assume a power law for the product, $n_{\mathrm{CR}} (r) \times n_{\mathrm{CGM}}(r) \propto r^{-a}$, where $a>0$ is a free parameter, $r\gg1$~kpc. We estimate $a$ from observations and assume that both the Milky Way and M31 have the same halo characteristic size and the same circumgalactic gas density profile shape $n_{\mathrm{CGM}}$, hence the knowledge of $a$ allows us to reconstruct $n_{\mathrm{CR}}$ up to its normalization. 
%The cosmic ray proton number density profile shape is derived from the source profile shape: $n_{\mathrm{CR}} \propto r^{-a} \times n_{\mathrm{CGM}}^{-1}$. 

\begin{figure}[tpb]
    \centering
    \includegraphics[width=\columnwidth]{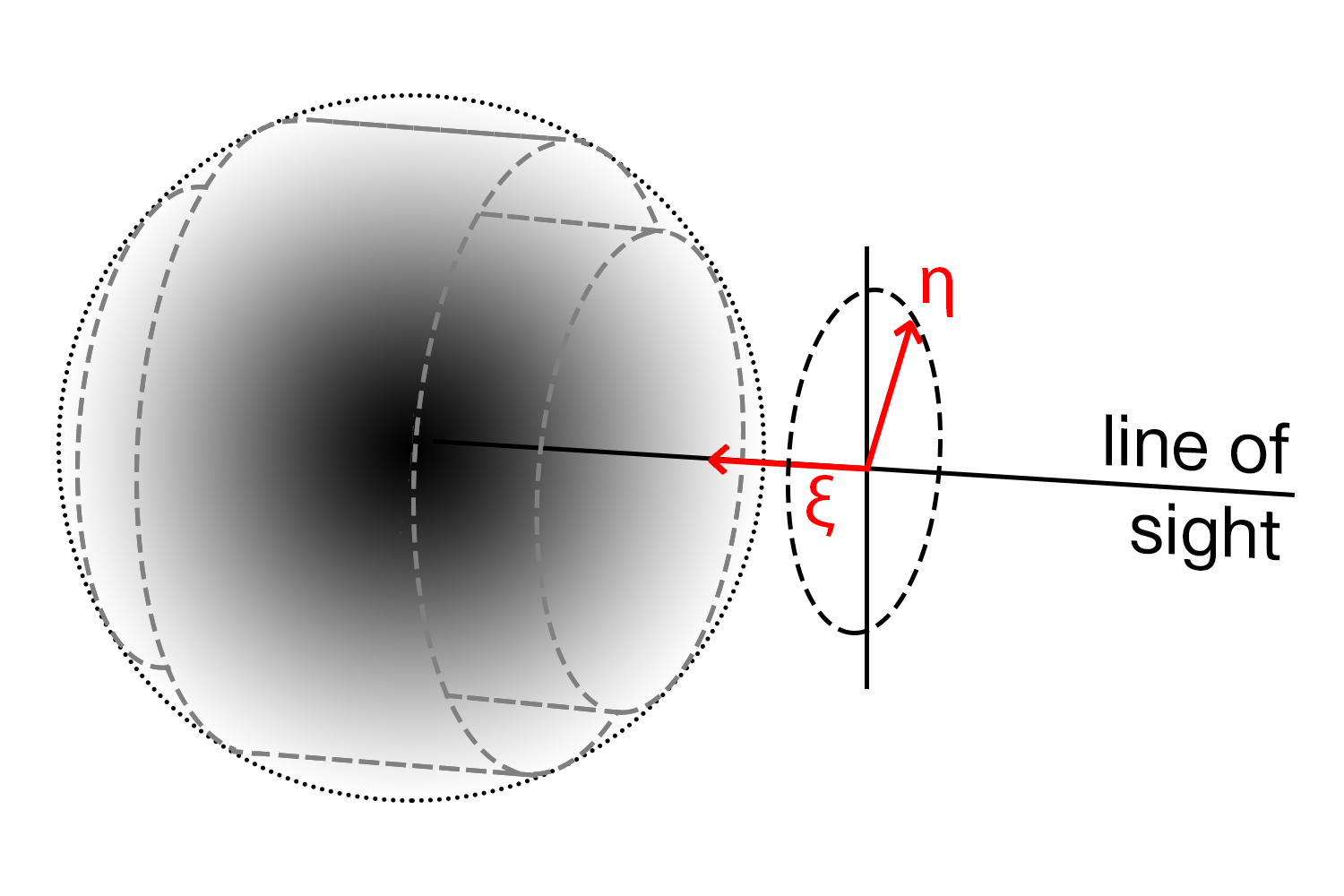}
    \caption{Schematic image of the M31 halo with the ring-shaped region contoured with dashed grey, see eq.~\eqref{eq:intensity}. For illustration, the origin is shifted from the galactic center along the line of sight.}
    \label{fig:M31halo}
\end{figure}

Consider a ring-shaped region determined in figure~\ref{fig:M31halo}. Its contribution to the observed intensity is
\begin{equation}
    \begin{gathered}
        I(\eta_{in}, \eta_{out} | a) = \frac{I_0}{\eta_{out}^2 - \eta_{in}^2} \int\limits_{\eta_{in}}^{\eta_{out}} d\eta\,\Bigg(\eta^{2}\times \int\limits_{0}^{\sqrt{1 - \eta^2}} d\xi\,\left(\xi^2 + \eta^2\right)^{-a/2}\Bigg) = \\ = \frac{I_0}{\eta_{out}^2 - \eta_{in}^2} \times \int\limits_{\eta_{in}}^{\eta_{out}} d\eta\,\eta^{2-a}\sqrt{1-\eta^2}~_2 F_1 \left(\frac{1}{2}, \frac{a}{2}, \frac{3}{2}, 1-\frac{1}{\eta^2}\right),
    \end{gathered}
    \label{eq:intensity}
\end{equation}
where $\eta$ and $\xi$ are the dimensionless coordinates perpendicular and along the line of sight (respectively), $\eta_{in}$ and $\eta_{out}$ determine the inner and outer projected radius of the region (respectively), and $_2 F_1$ is the hypergeometric function. 

By making use of eq.~\eqref{eq:intensity}, we calculate the intensity ratio $I_{\mathrm{SH}} \div I_{\mathrm{FOH}}$ and compare it with its observed value. The slope parameter $a=1.5$ corresponds to the observed ratio of 1.4 (hereafter referred to as the ``optimal'' profile), while $a=2.3$ gives the intensity ratio of $2.4$, which is the upper limit allowed by observations within the $1\sigma$ uncertainty (hereafter referred to as the ``sharp'' profile). The lower limit is not considered because it leads to $a<0$. 

\begin{figure}[tpb]
    \centering
    \includegraphics[width=\columnwidth]{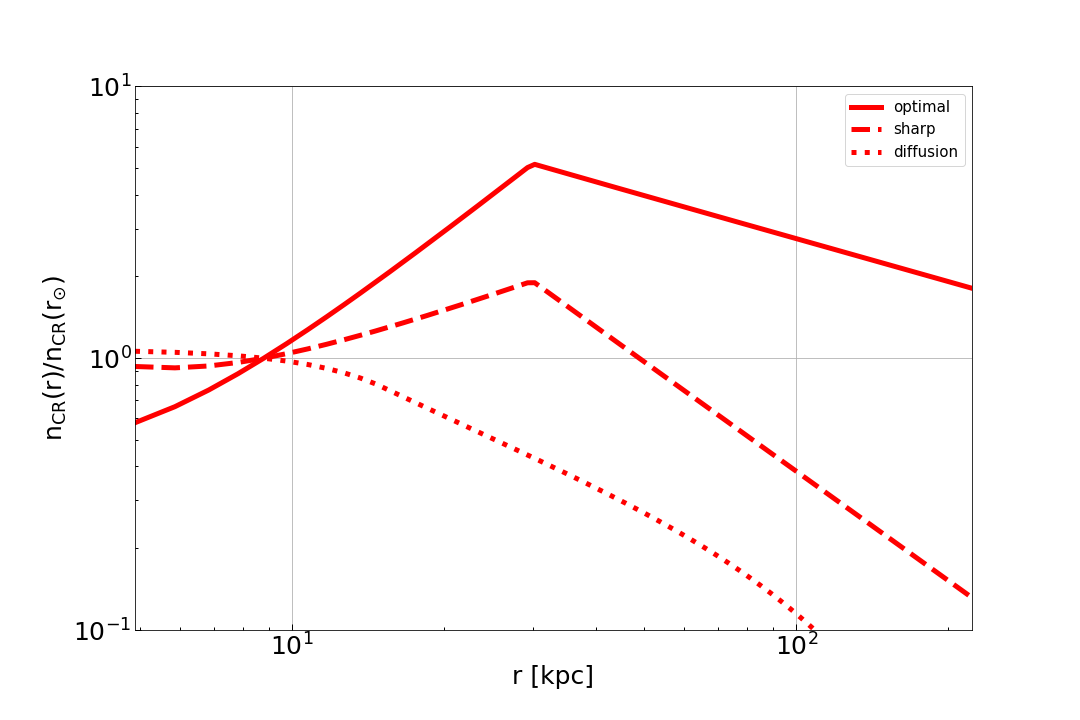}
    \caption{Cosmic-ray proton number density profile (normalized at $r_\odot=8.5$~kpc) for three models considered in this work.}
    \label{fig:CR-density:compare}
\end{figure}

 Since the diffusive and non-diffusive models have different injection spectra, the ratio of corresponding cosmic-ray spectral densities $j(E', r)$ depends on both radius and energy. In order to compare only the spatial distribution, we temporarily -- for illustration only -- use the energy-independent number density $n_{\rm{CR}}(r) = \int dE'~j(E', r)$ (where we integrate over $E' > 10$~GeV).
Figure~\ref{fig:CR-density:compare} compares the discussed non-diffusive cosmic-ray density profiles $n_{\rm{CR}}(r)$ with the previously obtained diffusive profile. 

Let us emphasize that the obtained profiles, ``optimal'' and ``sharp'', are applicable only within and, with caution, near the region in our Galaxy corresponding to the region between the inner and outer bounds of the M31 halo observed part (galactocentric radii of 5~kpc and 219~kpc, respectively). As the region of applicability is finite, the obtained $a$ parameter values do not imply that the total amount of cosmic rays associated with the Milky-Way (or the amount per spherical shell of constant thickness) diverges.

\section{Results and discussion}%(NM)
\label{sec:results}
\subsection{Circumgalactic neutrino flux}
\label{sec:res:flux}
\begin{figure}[H]
    \centering
    \includegraphics[width=\columnwidth]{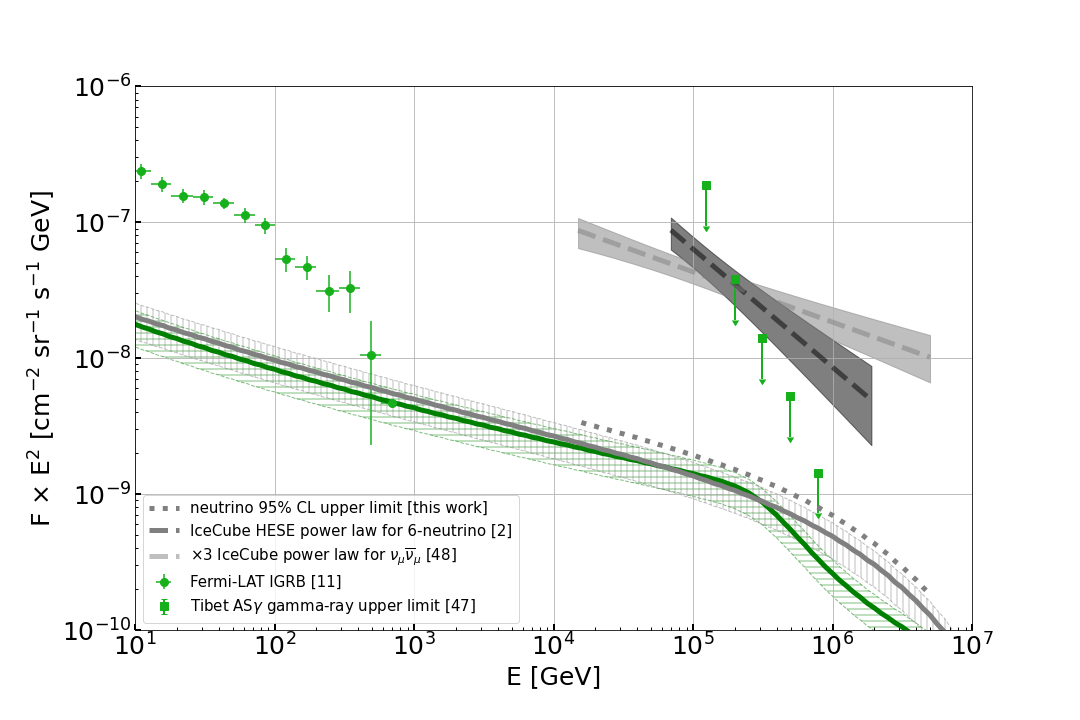}
    \includegraphics[width=\columnwidth]{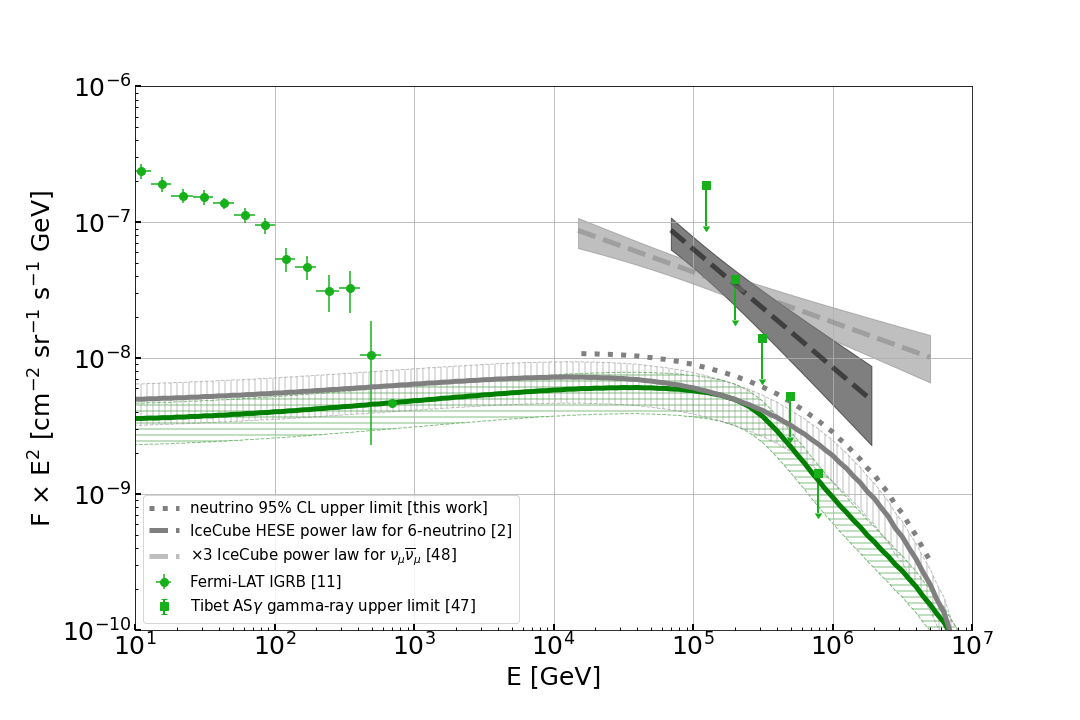}
    \caption{Gamma-ray (green) and 6-neutrino (grey) spectra. Solid lines and hatched regions correspond to the best-fit curves and 68\% confidence level uncertainty obtained in this work. Top: diffusive scenario, bottom: non-diffusive scenario. For comparison, observations from refs.~\cite{IceCube-HESE-2020,IceCube:muon} (neutrino) and~\cite{FermiDiffuse,TibetASgamma} (gamma rays) with their 68\% confidence level uncertainty are shown. See text for a detailed discussion.}
    \label{fig:ph+6nu_spectra}
\end{figure}

Figure~\ref{fig:ph+6nu_spectra} presents the resulting local gamma-ray and all-flavor neutrino fluxes calculated within the diffusive and the non-diffusive scenario and normalized using Fermi-LAT isotropic gamma-ray background (IGRB)~\cite{FermiDiffuse} (note that this background is model-dependent, here we use the results for Foreground model~A). Since the difference in these fluxes between the ``optimal'' and ``sharp'' model is negligible, the former model results are omitted. The flux uncertainty is due to the halo gas density profile uncertainty and is estimated as:

\begin{equation}
    \text{rel. err.} = \left(\int dr~\delta n_{\mathrm{CGM}} n_{\mathrm{CR}}\right) / \left(\int dr~n_{\mathrm{CGM}} n_{\mathrm{CR}}\right),
\end{equation}
where $\delta n_{\mathrm{CGM}}$ is the gas density profile uncertainty. The actual uncertainty may be larger due to the underestimated systematic errors in the assumed source spatial and spectral shape. However, since we use quite conservative assumptions on these shapes and normalize our results to local observations, we do not expect this underestimated uncertainty to change the results qualitatively. This level of precision is acceptable given the aim of the present work.

One can see that the total IceCube astrophysical neutrino flux cannot be explained by cosmic-ray interactions with the circumgalactic gas, both for diffusive and non-diffusive escape assumptions. Note that the circumgalactic neutrino flux at the energies of interest is $\simeq4$ times larger in the M31-inspired non-diffusive model than in the diffusive model. In any case, the associated gamma-ray flux agrees well with Fermi-LAT~\cite{FermiDiffuse} and Tibet AS$\gamma$~\cite{TibetASgamma} constraints. 

Table~\ref{tab:contribution} presents the fraction of the circumgalactic neutrinos in the total IceCube 6-neutrino flux. We define the total flux as $F^{\mathrm{tot}} = \int dE~F(E)$ [cm$^{-2}$ sr$^{-1}$ s$^{-1}$] and the fraction as $F^{\mathrm{tot}}_{\mathrm{modelled}}/F^{\mathrm{tot}}_{\mathrm{observed}}$. Note that this fraction not only depends on the escape scenario, but also on the assumptions about the IceCube spectrum. Here, we consider the ``single power-law'' models from ref.~\cite{IceCube-HESE-2020} (a single energy segment between 69.4~TeV and 1.9~PeV) and from ref.~\cite{IceCube:muon} (between 15.0~TeV and 5.0~PeV). For the latter model, we multiply the total observed flux by~3, since ref.~\cite{IceCube:muon} focuses only on $\nu_{\mu}\overline{\nu}_{\mu}$ flux.

\begin{figure}[tb]
    \centering
    \includegraphics[width=\columnwidth]{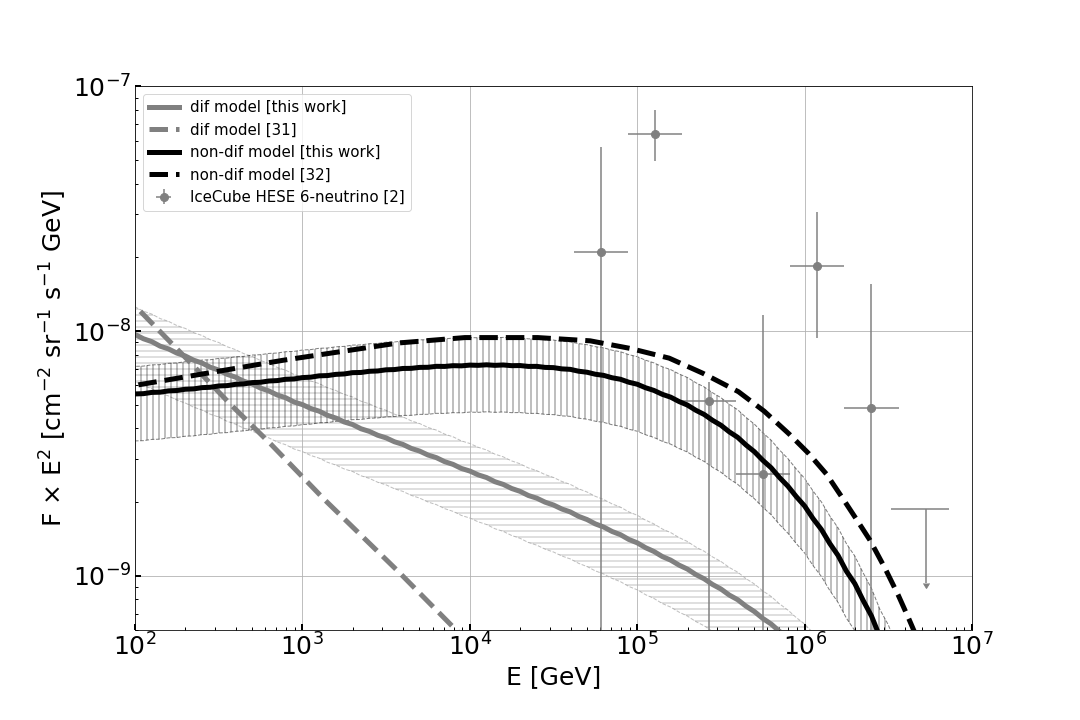}
    \caption{Neutrino fluxes for diffusive and non-diffusive scenario obtained in this work and in the most recent previous studies. Hatched regions correspond to 68\% confidence level uncertainty. For comparison, observations from ref.~\cite{IceCube-HESE-2020} (``segmented power-law'') are shown.}
    \label{fig:6nu_spectra_comparison}
\end{figure}

Figure~\ref{fig:6nu_spectra_comparison} compares our results with that of the two most recent papers on the same topic, refs.~\cite{KT2016} and~\cite{Gabici:2021-crhalo}. 

Within the diffusive scenario, we obtain a considerably larger neutrino flux compared to ref.~\cite{KT2016}, primarily because of assuming harder cosmic-ray spectrum. However, the modelled neutrino flux is still unlikely to exceed $\sim 4$\% of that observed. 

Within the non-diffusive scenario, we conclude the flux to be noticeably lower than that obtained in ref.~\cite{Gabici:2021-crhalo}. The difference is likely to be caused by the use of the refined circumgalactic-gas and cosmic-ray density profiles, since the assumed proton spectrum was in this work exactly the same as that in ref.~\cite{Gabici:2021-crhalo}. Our result does not support the idea that the circumgalactic neutrino flux could explain the total astrophysical neutrino flux observed by IceCube: we conclude that the flux is unlikely to exceed $\sim 17\%$ of that observed, and is $<50\%$ at the confidence level of more than $5\sigma$, not taking into account the systematic uncertainties discussed above. Moreover, if the opposite were true, that is if $\gtrsim 17\%$ of the neutrino flux were explained by this mechanism, the associated gamma-ray flux would be in a serious conflict with Fermi-LAT~\cite{FermiDiffuse} and Tibet-AS$\gamma$~\cite{TibetASgamma} constraints.

We also consider a normalization of the total integrated gamma-ray flux between $50$~GeV and $2$~TeV to Fermi-LAT extragalactic gamma-ray background (EGB) with the total integrated flux from point sources extracted~\cite{Fermi-LAT:2015otn}. In contrast to IGRB normalization, this approach does not suffer from foreground model dependency. The obtained gamma-ray and neutrino fluxes are approximately twice as large as those within IGRB normalization. However, this approach produces a considerable systematic uncertainty since the total integrated EGB without point sources' contribution is estimated as $(3.4\pm3.4)\times10^{-10}$~cm$^{-2}$~sr$^{-1}$~s$^{-1}$. Moreover, within the non-diffusive model, such a normalization leads to an overshoot of $\sim(2...3)\sigma$ of Tibet AS$\gamma$~\cite{TibetASgamma} constraints and the local proton spectrum observations~\cite{Dembinski:2017zsh,IceCube:2019hmk} at $E\sim10^{(5...6)}$~GeV and $E'\sim10^{(7...8)}$~GeV, respectively. Therefore, we hereafter discuss only the results normalized to Fermi-LAT IGRB~\cite{FermiDiffuse}.

\begin{table}[tpb]
    \centering
    \begin{tabular}{|l|c|c|}
        \hline
         \diagbox[width=12em]{model}{scenario} & diffusive & non-diffusive\\
         \hline
         IceCube HESE~\cite{IceCube-HESE-2020} & 2.7 (1.6...3.7)\% & 11.8 (6.8...16.6)\% \\
         \hline
         IceCube $\nu_{\mu}\overline{\nu}_{\mu}\times 3$~\cite{IceCube:muon} & 3.1 (1.9...4.1)\% & 11.1 (6.5...15.1)\% \\
         \hline
    \end{tabular}
    \caption{The fraction of the circumgalactic neutrinos in the total observed neutrino flux.  In brackets, 68\% confidence level intervals are shown.}
    \label{tab:contribution}
\end{table}

%\begin{figure}[htpb]
%    \centering
%    \includegraphics[width=\columnwidth]{}
%    \includegraphics[width=\columnwidth]{}
%    \caption{Gamma-ray (green) and 6-neutrino (grey) spectra. Solid lines and hatched regions correspond to the best-fit curves and 68\% confidence level uncertainty obtained in this work. Top: diffusive scenario, bottom: non-diffusive scenario. For comparison, observations from refs.~\cite{IceCube-HESE-2020,IceCube:muon} (neutrino) and~\cite{FermiDiffuse,TibetASgamma} (gamma rays) with their 68\% confidence level uncertainty are shown. See text for a detailed discussion.}
%    \label{fig:ph+6nu_spectra_EGB}
%\end{figure}

%\begin{table}[htpb]
%    \centering
%    \begin{tabular}{|l|c|c|}
%        \hline
%         \diagbox[width=10em]{model}{scenario} & diffusive & non-diffusive\\
%         \hline
%         IceCube HESE~\cite{IceCube-HESE-2020} & 5.1 (3.1...7.0)\% & 45.1 (25.9...63.2)\% \\
%         \hline
%         IceCube $\nu_{\mu}\overline{\nu}_{\mu}\times 3$~\cite{IceCube:muon} & 5.9 (3.7...7.9)\% & 42.1 (24.9...57.5)\% \\
%         \hline
%    \end{tabular}
%    \caption{The fraction of the circumgalactic neutrinos in the total observed neutrino flux.  In brackets, 68\% confidence level intervals are shown. Note that the systematic uncertainty from the normalization is not included, and the non-diffusive model results are in conflict with observational constraints~\cite{TibetASgamma,Dembinski:2017zsh,IceCube:2019hmk}. See text for a detailed discussion.}
%    \label{tab:contribution_EGB}
%\end{table}

\subsection{Energetics}
\label{sec:res:energy}
One can estimate the total energy of cosmic rays within the virial radius of the Galaxy as $E'_{\rm{tot}} = \int dE'\int 4\pi r^2dr~E'j(E', r)$, where we integrate over $E'>10$~GeV and over the Milky-Way halo volume, $r<R_{\rm{vir}}=223$~kpc (see ref.~\cite{2020SCPMA..6309801W}). This energy can be considered as a probe of the cosmic-ray source power required to produce the corresponding cosmic rays distribution. If $T$ is the characteristic period of the source activity, the characteristic power $P$ is constrained as $P>E'_{\rm{tot}}/T$. Table~\ref{tab:energetics} briefly compares the diffusive and non-diffusive (``sharp'' model) scenarios in terms of their energetics. 

\begin{table}[tpb]
    \centering
    \begin{tabular}{|l|c|c|}
        \hline
         \diagbox[width=14em]{parameter}{scenario} & diffusive & non-diffusive\\
         \hline
         $E'_{\rm{tot}}$, $10^{55}$~erg & 2.9 & 3.1 \\
         \hline
         $P$, $10^{41}$~erg~s$^{-1}\times(T/\text{Myr})$ & 9.3 & 9.8 \\
         \hline
    \end{tabular}
    \caption{Cosmic-ray energetics within the diffusive and non-diffusive scenarios.}
    \label{tab:energetics}
\end{table}

Interestingly, there is no qualitative difference in terms of the required source power: for both scenarios, the lower limit is $P\times(T/\text{Myr})\sim10^{42}$~erg~s$^{-1}$ (for the ``optimal'' non-diffusive model, this limit is $\sim 10$ times larger). This constraint agrees with the hypothesis that the required power is associated with the Galactic-center activity, which reveals itself in Fermi Bubbles (see refs.~\cite{FB:2012a,FB:2012b,FB:2014,FB:2016}, where the total injected energies of $E_{\rm{FB}}\simeq10^{(55...58)}$~erg and the injection rates of $P_{\rm{FB}}\simeq10^{(41...43)}$~erg~s$^{-1}$ are reported).

We conclude that the larger neutrino flux obtained in the non-diffusive escape model, compared to that for the diffusive case, is due to a combination of the hard proton spectrum (the initial $E'^{-2}$ spectrum assumed in both cases softens in the diffusive scenario) and the specific cosmic-ray radial distribution shape. The latter is supported by the fact that the non-diffusive cosmic-ray concentration profile peaks at $r\simeq30$~kpc, i.e.\ the circumgalactic neutrinos are effectively collected from the larger volume, compared to the diffusive scenario.
This peak position fits well the idea that the Galactic-center activity might generate cosmic rays propagating to the Milky-Way halo without contributing significantly to the locally observed cosmic-ray energy density, see refs.~\cite{Gabici:2021-crhalo,FB:2012a}. It should be however remembered that in the present work, the cosmic-ray profile is determined from M31 observations and not from simulations; the peak appears due to the assumed shape of $n_{\rm{CGM}}^{-1}$ (see section~\ref{sec:anal:ah}). We defer the discussion of possible mechanisms of such propagation for further studies.

\subsection{Extragalactic contribution}
\label{sec:res:EG}
Neutrinos propagate through the Universe without absorption, hence similar interactions in circumgalactic gas halos of other galaxies also contribute to the locally observed neutrino flux. Within our model, the intensity of a source corresponding to a single external galaxy is
\begin{equation}
    I_{\mathrm{gal}} = \mathrm{const} \times \int\limits_{0}^{R_{\mathrm{vir,~gal}}} dr~n_{\mathrm{CGM}} n_{\mathrm{CR}} 4\pi r^2.
    \label{eq:eg_intensity}
\end{equation}
For simplicity, we hereafter assume that this intensity depends only on the galactic stellar mass $M^{\star}$, namely $I_{\mathrm{gal}} / I_{\mathrm{MW}} = \left(M^{\star}_{\mathrm{gal}}/M^{\star}_{\mathrm{MW}}\right)^\eta$, where $\eta\sim1$ is a free parameter. The corresponding neutrino flux produced by this galaxy is
\begin{equation}
    F_{\mathrm{gal}} = \frac{I_{\mathrm{gal}}}{4\pi d_{L, \mathrm{gal}}^2},
\end{equation}
where $d_L = 
%(1+z)d = 
(1+z) c \int_{0}^{z} dz'~H(z')^{-1}$ is the luminosity distance, $z$ is the redshift, $H$ is the Hubble constant. The Milky-Way-associated flux can be estimated as the flux from the Galactic anti-center,
\begin{equation}
    F_{\mathrm{MW}} = \mathrm{const} \times \int\limits_{r_{\odot}}^{R_{\mathrm{vir,~MW}}} dr~n_{\mathrm{CGM}} n_{\mathrm{CR}},
\end{equation}
where ``const'' is the same constant as that in eq.~\eqref{eq:eg_intensity}. The ratio between the extragalactic and Milky-Way-associated flux is thus:
\begin{equation}
    \begin{gathered}
         \zeta \equiv F_{\mathrm{EG}}/F_{\mathrm{MW}} = \sum_{\mathrm{gal}}~F_{\mathrm{gal}}/F_{\mathrm{MW}} =%\\= \int dV~\frac{n_{\mathrm{gal}} (z)}{4\pi d^2(z) (1+z)^2} \left(M^{\star}_{\mathrm{gal}} (z) /M^{\star}_{\mathrm{MW}}\right)^\eta\times\\\times\left(\frac{\int\limits_{0}^{R_{\mathrm{vir,~MW}}} dr~n_{\mathrm{CGM}} n_{\mathrm{CR}} 4\pi r^2}{\int\limits_{r_{\odot}}^{R_{\mathrm{vir,~MW}}} dr~n_{\mathrm{CGM}} n_{\mathrm{CR}}}\right) =
         \\= \frac{c}{H_0}\int dz~\frac{n^{1-\eta}_{\mathrm{gal}} (z) (1+z)^{-2}}{\sqrt{\Omega_{\Lambda} + \Omega_{M}(1+z)^3}} \left(\rho^{\star}_{\mathrm{gal}} (z) /M^{\star}_{\mathrm{MW}}\right)^\eta 
         \times\left(\frac{\int\limits_{0}^{R_{\mathrm{vir,~MW}}} dr~n_{\mathrm{CGM}} n_{\mathrm{CR}} 4\pi r^2}{\int\limits_{r_{\odot}}^{R_{\mathrm{vir,~MW}}} dr~n_{\mathrm{CGM}} n_{\mathrm{CR}}}\right)
    \end{gathered}
\end{equation}
In this equation, $\sum_{\mathrm{gal}}$ denotes the sum over all external galaxies, $n_{\mathrm{gal}},~\rho^{\star}_{\mathrm{gal}}$ are the galaxies' comoving number density and the stellar comoving mass density, $\Omega_{\Lambda, M}$ and $H_0$ are the cosmological parameters. Following the approach of ref.~\cite{KT2016}, we obtain 
%$\rho^{\star}_{\mathrm{gal}}$ as follows:
\begin{equation}
    \rho^{\star}_{\mathrm{gal}} = \int \Phi M^{\star} d\log M^{\star},\qquad\Phi = \frac{dN}{dV d\log M^{\star}}.
\end{equation}
To obtain $\zeta(\eta)$, we use $\Phi(z)$ and $n_{\mathrm{gal}}(z)$ presented in refs.~\cite{2015MNRAS.450.1604L} and~\cite{2015A&A...583A..61G}, respectively. 

Although the result depends considerably on the assumed value of $\eta$, in all realistic cases one finds $\zeta<0.1$ (for the both considered scenarios; the larger $\eta$, the smaller $\zeta$) and thus the extragalactic neutrino flux from other halos does not increase the one associated with the Milky Way beyond the accuracy of our flux estimates. Therefore, we conclude the extragalactic contribution to be negligible within our analysis.

\section{Conclusions}
\label{sec:concl}
In this work, we estimate the flux of neutrinos born in interactions of cosmic rays, leaving the Milky-Way disk, with the circumgalactic gas, for two alternative scenarios of cosmic-ray escape. 

One scenario assumes that cosmic rays escape diffusively from the Galactic center towards the outer regions of the Milky-Way halo. This study set out to revise findings presented in ref.~\cite{KT2016} using the updated circumgalactic gas density profile~\cite{Martynenko:2021} and assuming the cosmic-ray source to have a hard spectrum ($E'^{-2}$) and to fill a large volume near the Milky-Way disk ($r<15$~kpc). In this case, we find the circumgalactic neutrino flux to contribute $\sim (1.6...3.7)\%$ of that observed by IceCube (at $68\%$ confidence level). Despite this is not enough to explain IceCube's observations, the flux is comparable with the Galactic-disk contribution $\sim (4...8)\%$ (see e.g.\ ref.~\cite{2016:GalNu}). 

For the second scenario, the assumption of the diffusive escape is lifted. Instead of modelling the cosmic-ray number density profile around the Milky Way, we assume, following ref.~\cite{Gabici:2021-crhalo}, a similarity between our Galaxy and M31, and tune the cosmic-ray profile to reproduce gamma-ray observations of the M31 circumgalactic environment (without specifying the mechanism of cosmic-ray propagation). We find that the circumgalactic neutrino flux could contribute $\sim(6.8...16.6)\%$ of the observed IceCube flux in this case (at $68\%$ confidence level). We also find this contribution to be $<50\%$ at the confidence level of more than $5\sigma$, and argue that the contribution of $\gtrsim 17\%$ leads to a conflict with gamma-ray observations. The latter finding disagrees with the idea that the dominant contributor to IceCube's observations may be the Milky-Way circumgalactic medium. A note of caution is due here since the M31 observations correspond to the photon energy range of $E<100$~GeV (while the range of interest is near $\sim100$~TeV), and the measurements are characterized by a large degree of uncertainty.

Particular quantitative results may vary considerably depending on the assumptions about poorly constrained cosmic-ray profile, as well as the additional non-spherical component of both neutrinos and gamma-rays produced in the Galactic halo substructures can enlarge the total flux, so the estimated neutrino fluxes should be considered as ballpark values only. However, these variations in the assumptions could not change our main qualitative conclusion: the contribution of cosmic-ray interactions in the circumgalactic gas cannot explain the entire flux of astrophysical neutrinos observed by IceCube. This is because, as discussed in section~\ref{sec:res:flux}, higher neutrino fluxes would contradict diffuse gamma-ray measurements at the Earth.

Note that in both cases, the energetic requirements are of the same order of magnitude and do not exceed the approximate energetic capabilities of the Galactic center: the total energy held in cosmic rays is $E'_{\rm{tot}}\sim3\times10^{55}$~erg, and the corresponding source characteristic power is $P\times(T/\text{Myr})\sim10^{42}$~erg~s$^{-1}$. 

We also estimate the neutrino flux from external galaxies' halos and find that this is further suppressed with respect to the Galactic contribution in all realistic cases and is negligible within the accuracy of our analysis.

\appendix
\section{Numerical solution}
\label{app:numsol}
In order to solve eq.~\eqref{eq:diffusion}, we define an auxiliary function $u(E', r, t) = r j(E', r, t)/Q_0$, where $Q_0$ is the normalization of the source term $Q_{\mathrm{p}} (E', r, t)$. Then eq.~\eqref{eq:diffusion} leads to:
\begin{equation}
    \begin{gathered}
        \left[\partial_t - D(E')\partial^2_r + c\sigma_{\mathrm{pp}}(E')n_{\mathrm{CGM}}(r)\right]u(E', r, t) = \\
        = E'^{-\alpha} \exp\left(\frac{-E'}{E'_{\mathrm{cut}}}\right) \times r\theta(r_{\mathrm{Q}} - r) \times Q_{\mathrm{time}}(t)
    \end{gathered}
\end{equation}
Since we assume a spherical symmetry, we expect $j(E', r, t) = j(E', -r, t)$ and therefore $\partial_r j(E', r=0, t) = 0$. In addition, we assume $j(E', r \rightarrow \infty, t) = 0$ due to the fact that the proton propagation is limited to a finite region which size is determined by the corresponding gyroradius. We adopt a zero initial condition $j(E', r, t=0) = 0$ considering only the cosmic rays produced after the final assembly of the inner Galactic halo.

Let us fix $E'$ and denote $r_{\text{kpc}}\equiv\varrho$, $t_{\text{Gyr}}\equiv\tau$, $D(E')\times\text{kpc}^{-2}\text{~Gyr}\equiv D$, $c\sigma_{\mathrm{pp}}(E')n_{\mathrm{CGM}}(r) \times \text{Gyr} \equiv f(\varrho)$, $E'^{-\alpha} \exp\left(\frac{-E'}{E'_{\mathrm{cut}}}\right)r\theta(r_{\mathrm{Q}} - r) Q_{\mathrm{time}}(t) \times \text{GeV}^{\alpha}\text{ kpc}^{-1} \equiv q(\varrho, \tau)$.

We construct a two-dimensional rectangular grid $(\varrho_k, \tau_m) = (k\Delta\varrho, m\Delta\tau)$ on the region of $(\varrho, \tau) \in [0, \varrho_{\mathrm{bound}}] \times [0, 10]$. For calculations within the ``simple'' (``realistic'') model of source evolution we use $1024\times256$ ($1024\times2048$) grid with $\varrho_{\rm{bound}}=1000$. The latter corresponds to the physical radius of 1~Mpc, which is several orders of magnitude larger than a characteristic gyroradius for a proton at the energy of $E'_{\rm{cut}}$ in the circumgalactic magnetic field with a typical strength of $\sim 10^{-(1...2)} \mu\text{G}$ (see ref.~\cite{2020MNRAS.498.3125P}). On this grid, we adopt the following unconditionally stable difference scheme:
\begin{equation}
         \frac{u^{m+1}_{k}-u^{m}_{k}}{\Delta\tau} = \frac{D\left(u^{m+1}_{k+1} - u^{m+1}_{k-1} - 2u^{m+1}_{k}\right)}{(\Delta \rho)^2} - f_{k} u^{m+1}_{i} + q^{m+1}_{k},
\end{equation}
where the indices $m$ and $k$ correspond to the coordinates of $\rho_m$ and $\tau_k$, respectively. In terms of $u^{m}_{k}$, the discussed boundary and initial conditions lead to $u^{0}_{k} = u^{m}_{0} = u^{m}_{k_{\mathrm{max}}}$. We find $u^{m}_{k}$ using the tridiagonal matrix algorithm.

\acknowledgments
We are indebted to Felix Aharonian, Sarah Recchia, Dmitri Semikoz and Andrew Taylor for interesting discussions. This work is supported by the RF Ministry of science and higher education under the contract 075-15-2020-778. NM thanks the Theoretical Physics and Mathematics Advancement Foundation ``BASIS'' for the fellowship under the contract 21-2-1-65-1.

\bibliography{circumnu}

\providecommand{\href}[2]{#2}\begingroup\raggedright\begin{thebibliography}{10}

\bibitem{IceCube-2013}
{\scshape IceCube} collaboration, \emph{{Evidence for High-Energy
  Extraterrestrial Neutrinos at the IceCube Detector}},
  \href{https://doi.org/10.1126/science.1242856}{\emph{Science} {\bfseries 342}
  (2013) 1242856} [\href{https://arxiv.org/abs/1311.5238}{{\ttfamily
  1311.5238}}].

\bibitem{IceCube-HESE-2020}
{\scshape IceCube} collaboration, \emph{{The IceCube high-energy starting event
  sample: Description and flux characterization with 7.5 years of data}},
  \href{https://doi.org/10.1103/PhysRevD.104.022002}{\emph{Phys. Rev. D}
  {\bfseries 104} (2021) 022002}
  [\href{https://arxiv.org/abs/2011.03545}{{\ttfamily 2011.03545}}].

\bibitem{ANTARES-diffuse}
{\scshape ANTARES} collaboration, \emph{{Study of the high-energy neutrino
  diffuse flux with the ANTARES neutrino telescope}},
  \href{https://doi.org/10.22323/1.358.0891}{\emph{PoS} {\bfseries ICRC2019}
  (2020) 891}.

\bibitem{Baikal-Neutrino2022a}
{\scshape Baikal-GVD} collaboration, Z.~Dzhilkibaev, ``{Status of the
  Baikal-GVD and selected results}.'' {Talk at the XXX International conference
  on neutrino physics and astrophysics (Neutrino-2022), Seoul (virtual), May
  29--June 4}, 2022.

\bibitem{Baikal:2022chp}
{\scshape Baikal} collaboration, \emph{{Diffuse neutrino flux measurements with
  the Baikal-GVD neutrino telescope}},
  \href{https://arxiv.org/abs/2211.09447}{{\ttfamily 2211.09447}}.

\bibitem{ST-UFN}
S.~Troitsky, \emph{{Constraints on models of the origin of high-energy
  astrophysical neutrinos}},
  \href{https://doi.org/10.3367/UFNe.2021.09.039062}{\emph{Usp. Fiz. Nauk}
  {\bfseries 191} (2021) 1333}
  [\href{https://arxiv.org/abs/2112.09611}{{\ttfamily 2112.09611}}].

\bibitem{IceCubeANTARES-GalPlane}
{\scshape ANTARES, IceCube} collaboration, \emph{{Joint Constraints on Galactic
  Diffuse Neutrino Emission from the ANTARES and IceCube Neutrino Telescopes}},
  \href{https://doi.org/10.3847/2041-8213/aaeecf}{\emph{Astrophys. J. Lett.}
  {\bfseries 868} (2018) L20}
  [\href{https://arxiv.org/abs/1808.03531}{{\ttfamily 1808.03531}}].

\bibitem{Nikishov1962}
A.I.~{Nikishov}, \emph{{Absorption of High-Energy Photons in the Universe}},
  {\emph{JETP} {\bfseries 14} (1962) 393}.

\bibitem{BerezinskyKalashev}
V.~Berezinsky and O.~Kalashev, \emph{{High energy electromagnetic cascades in
  extragalactic space: physics and features}},
  \href{https://doi.org/10.1103/PhysRevD.94.023007}{\emph{Phys. Rev. D}
  {\bfseries 94} (2016) 023007}
  [\href{https://arxiv.org/abs/1603.03989}{{\ttfamily 1603.03989}}].

\bibitem{BerezinskySmirnov1975}
V.S.~Berezinsky and A.Y.~Smirnov, \emph{{Cosmic neutrinos of ultra-high
  energies and detection possibility}},
  \href{https://doi.org/10.1007/BF00643157}{\emph{Astrophys. Space Sci.}
  {\bfseries 32} (1975) 461}.

\bibitem{FermiDiffuse}
{\scshape Fermi-LAT} collaboration, \emph{{The spectrum of isotropic diffuse
  gamma-ray emission between 100 MeV and 820 GeV}},
  \href{https://doi.org/10.1088/0004-637X/799/1/86}{\emph{Astrophys. J.}
  {\bfseries 799} (2015) 86} [\href{https://arxiv.org/abs/1410.3696}{{\ttfamily
  1410.3696}}].

\bibitem{MuraseCorona1}
K.~Murase, S.S.~Kimura and P.~Meszaros, \emph{{Hidden Cores of Active Galactic
  Nuclei as the Origin of Medium-Energy Neutrinos: Critical Tests with the MeV
  Gamma-Ray Connection}},
  \href{https://doi.org/10.1103/PhysRevLett.125.011101}{\emph{Phys. Rev. Lett.}
  {\bfseries 125} (2020) 011101}
  [\href{https://arxiv.org/abs/1904.04226}{{\ttfamily 1904.04226}}].

\bibitem{AhlersMurase-gamma-Gal}
M.~Ahlers and K.~Murase, \emph{{Probing the Galactic Origin of the IceCube
  Excess with Gamma-Rays}},
  \href{https://doi.org/10.1103/PhysRevD.90.023010}{\emph{Phys. Rev. D}
  {\bfseries 90} (2014) 023010}
  [\href{https://arxiv.org/abs/1309.4077}{{\ttfamily 1309.4077}}].

\bibitem{KalashevST-gamma-Gal}
O.E.~Kalashev and S.V.~Troitsky, \emph{{IceCube astrophysical neutrinos without
  a spectral cutoff and $10^{15}$\textendash{}$10^{17}$ eV cosmic gamma
  radiation}}, \href{https://doi.org/10.1134/S0021364014240072}{\emph{Pisma Zh.
  Eksp. Teor. Fiz.} {\bfseries 100} (2014) 865}
  [\href{https://arxiv.org/abs/1410.2600}{{\ttfamily 1410.2600}}].

\bibitem{2comp-Chen}
C.-Y.~Chen, P.S.~Bhupal~Dev and A.~Soni, \emph{{Two-component flux explanation
  for the high energy neutrino events at IceCube}},
  \href{https://doi.org/10.1103/PhysRevD.92.073001}{\emph{Phys. Rev. D}
  {\bfseries 92} (2015) 073001}
  [\href{https://arxiv.org/abs/1411.5658}{{\ttfamily 1411.5658}}].

\bibitem{2comp-Vissani}
A.~Palladino and F.~Vissani, \emph{{Extragalactic plus Galactic model for
  IceCube neutrino events}},
  \href{https://doi.org/10.3847/0004-637X/826/2/185}{\emph{Astrophys. J.}
  {\bfseries 826} (2016) 185}
  [\href{https://arxiv.org/abs/1601.06678}{{\ttfamily 1601.06678}}].

\bibitem{2comp-Neronov}
A.~Neronov and D.~Semikoz, \emph{{Galactic and extragalactic contributions to
  the astrophysical muon neutrino signal}},
  \href{https://doi.org/10.1103/PhysRevD.93.123002}{\emph{Phys. Rev. D}
  {\bfseries 93} (2016) 123002}
  [\href{https://arxiv.org/abs/1603.06733}{{\ttfamily 1603.06733}}].

\bibitem{2comp-Vissani-2}
A.~Palladino, M.~Spurio and F.~Vissani, \emph{{On the IceCube spectral
  anomaly}}, \href{https://doi.org/10.1088/1475-7516/2016/12/045}{\emph{JCAP}
  {\bfseries 12} (2016) 045}
  [\href{https://arxiv.org/abs/1610.07015}{{\ttfamily 1610.07015}}].

\bibitem{neutradio1}
A.~Plavin, Y.Y.~Kovalev, Y.A.~Kovalev and S.~Troitsky, \emph{{Observational
  Evidence for the Origin of High-energy Neutrinos in Parsec-scale Nuclei of
  Radio-bright Active Galaxies}},
  \href{https://doi.org/10.3847/1538-4357/ab86bd}{\emph{Astrophys. J.}
  {\bfseries 894} (2020) 101}
  [\href{https://arxiv.org/abs/2001.00930}{{\ttfamily 2001.00930}}].

\bibitem{hovatta}
T.~Hovatta et~al., \emph{{Association of IceCube neutrinos with radio sources
  observed at Owens Valley and Mets\"ahovi Radio Observatories}},
  \href{https://doi.org/10.1051/0004-6361/202039481}{\emph{Astron. Astrophys.}
  {\bfseries 650} (2021) A83}
  [\href{https://arxiv.org/abs/2009.10523}{{\ttfamily 2009.10523}}].

\bibitem{neutradio2}
A.V.~Plavin, Y.Y.~Kovalev, Y.A.~Kovalev and S.V.~Troitsky, \emph{{Directional
  Association of TeV to PeV Astrophysical Neutrinos with Radio Blazars}},
  \href{https://doi.org/10.3847/1538-4357/abceb8}{\emph{Astrophys. J.}
  {\bfseries 908} (2021) 157}
  [\href{https://arxiv.org/abs/2009.08914}{{\ttfamily 2009.08914}}].

\bibitem{ANTARES-radioJulien}
{\scshape ANTARES} collaboration, \emph{{Search for an association between
  neutrinos and radio-selected blazars with ANTARES}},
  \href{https://doi.org/10.22323/1.395.1164}{\emph{PoS} {\bfseries ICRC2021}
  (2021) 1164}.

\bibitem{ANTARES-radioGiulia}
{\scshape ANTARES} collaboration, \emph{{ANTARES search for neutrino flares
  from the direction of radio-bright blazars}},
  \href{https://doi.org/10.22323/1.395.0972}{\emph{PoS} {\bfseries ICRC2021}
  (2021) 972}.

\bibitem{neutGalaxy}
Y.Y.~Kovalev, A.V.~Plavin and S.V.~Troitsky, \emph{{Galactic Contribution to
  the High-energy Neutrino Flux Found in Track-like IceCube Events}},
  \href{https://doi.org/10.3847/2041-8213/aca1ae}{\emph{Astrophys. J. Lett.}
  {\bfseries 940} (2022) L41}
  [\href{https://arxiv.org/abs/2208.08423}{{\ttfamily 2208.08423}}].

\bibitem{ANTARES-GalRidge}
{\scshape ANTARES} collaboration, \emph{{Hint for a TeV neutrino emission from
  the Galactic Ridge with ANTARES}},
  \href{https://arxiv.org/abs/2212.11876}{{\ttfamily 2212.11876}}.

\bibitem{NeronovBubble2018}
A.~Neronov, M.~Kachelrie\ss{} and D.V.~Semikoz, \emph{{Multimessenger gamma-ray
  counterpart of the IceCube neutrino signal}},
  \href{https://doi.org/10.1103/PhysRevD.98.023004}{\emph{Phys. Rev. D}
  {\bfseries 98} (2018) 023004}
  [\href{https://arxiv.org/abs/1802.09983}{{\ttfamily 1802.09983}}].

\bibitem{LocalBubble2020}
M.~Bouyahiaoui, M.~Kachelrie\ss{} and D.V.~Semikoz, \emph{{High-energy
  neutrinos from cosmic ray interactions in the Local Bubble}},
  \href{https://doi.org/10.1103/PhysRevD.101.123023}{\emph{Phys. Rev. D}
  {\bfseries 101} (2020) 123023}
  [\href{https://arxiv.org/abs/2001.00768}{{\ttfamily 2001.00768}}].

\bibitem{Taylor:halo-neutrino}
A.M.~Taylor, S.~Gabici and F.~Aharonian, \emph{{Galactic halo origin of the
  neutrinos detected by IceCube}},
  \href{https://doi.org/10.1103/PhysRevD.89.103003}{\emph{Phys. Rev. D}
  {\bfseries 89} (2014) 103003}
  [\href{https://arxiv.org/abs/1403.3206}{{\ttfamily 1403.3206}}].

\bibitem{Gupta:2012-gas}
A.~Gupta, S.~Mathur, Y.~Krongold, F.~Nicastro and M.~Galeazzi, \emph{{A huge
  reservoir of ionized gas around the Milky Way: Accounting for the Missing
  Mass?}}, \href{https://doi.org/10.1088/2041-8205/756/1/L8}{\emph{Astrophys.
  J. Lett.} {\bfseries 756} (2012) L8}
  [\href{https://arxiv.org/abs/1205.5037}{{\ttfamily 1205.5037}}].

\bibitem{Feldmann:halo-photons}
R.~Feldmann, D.~Hooper and N.Y.~Gnedin, \emph{{Circum-Galactic Gas and the
  Isotropic Gamma Ray Background}},
  \href{https://doi.org/10.1088/0004-637X/763/1/21}{\emph{Astrophys. J.}
  {\bfseries 763} (2013) 21} [\href{https://arxiv.org/abs/1205.0249}{{\ttfamily
  1205.0249}}].

\bibitem{KT2016}
O.~Kalashev and S.~Troitsky, \emph{{Fluxes of diffuse gamma rays and neutrinos
  from cosmic-ray interactions with the circumgalactic gas}},
  \href{https://doi.org/10.1103/PhysRevD.94.063013}{\emph{Phys. Rev. D}
  {\bfseries 94} (2016) 063013}
  [\href{https://arxiv.org/abs/1608.07421}{{\ttfamily 1608.07421}}].

\bibitem{Gabici:2021-crhalo}
S.~Gabici, S.~Recchia, F.~Aharonian and V.~Niro, \emph{{Giant Cosmic-Ray Halos
  around M31 and the Milky Way}},
  \href{https://doi.org/10.3847/1538-4357/abfda4}{\emph{Astrophys. J.}
  {\bfseries 914} (2021) 135}
  [\href{https://arxiv.org/abs/2101.05016}{{\ttfamily 2101.05016}}].

\bibitem{Kalashev:2014xna}
O.E.~Kalashev and E.~Kido, \emph{{Simulations of Ultra High Energy Cosmic Rays
  propagation}}, \href{https://doi.org/10.1134/S1063776115040056}{\emph{J. Exp.
  Theor. Phys.} {\bfseries 120} (2015) 790}
  [\href{https://arxiv.org/abs/1406.0735}{{\ttfamily 1406.0735}}].

\bibitem{2010ApJ...710.1530V}
T.M.~{Venters}, \emph{{Contribution to the Extragalactic Gamma-Ray Background
  from the Cascades of very High Energy Gamma Rays from Blazars}},
  \href{https://doi.org/10.1088/0004-637X/710/2/1530}{\emph{Astrophys. J.}
  {\bfseries 710} (2010) 1530}
  [\href{https://arxiv.org/abs/1001.1363}{{\ttfamily 1001.1363}}].

\bibitem{2016PhRvD..94b3007B}
V.~{Berezinsky} and O.~{Kalashev}, \emph{{High-energy electromagnetic cascades
  in extragalactic space: Physics and features}},
  \href{https://doi.org/10.1103/PhysRevD.94.023007}{\emph{Phys. Rev. D}
  {\bfseries 94} (2016) 023007}
  [\href{https://arxiv.org/abs/1603.03989}{{\ttfamily 1603.03989}}].

\bibitem{Fermi-LAT:2015otn}
{\scshape Fermi-LAT} collaboration, \emph{{Resolving the Extragalactic
  $\gamma$-Ray Background above 50 GeV with the Fermi Large Area Telescope}},
  \href{https://doi.org/10.1103/PhysRevLett.116.151105}{\emph{Phys. Rev. Lett.}
  {\bfseries 116} (2016) 151105}
  [\href{https://arxiv.org/abs/1511.00693}{{\ttfamily 1511.00693}}].

\bibitem{2018ApJ...862....3B}
J.N.~{Bregman}, M.E.~{Anderson}, M.J.~{Miller}, E.~{Hodges-Kluck}, X.~{Dai},
  J.-T.~{Li} et~al., \emph{{The Extended Distribution of Baryons around
  Galaxies}}, \href{https://doi.org/10.3847/1538-4357/aacafe}{\emph{Astrophys.
  J.} {\bfseries 862} (2018) 3}
  [\href{https://arxiv.org/abs/1803.08963}{{\ttfamily 1803.08963}}].

\bibitem{Martynenko:2021}
N.~Martynenko, \emph{{Constraining density and metallicity of the Milky
  Way\textquoteright{}s hot gas halo from O\,vii spectra and ram-pressure
  stripping}}, \href{https://doi.org/10.1093/mnras/stac164}{\emph{Mon. Not.
  Roy. Astron. Soc.} {\bfseries 511} (2022) 843}
  [\href{https://arxiv.org/abs/2105.02557}{{\ttfamily 2105.02557}}].

\bibitem{2016A&A...592A.142R}
T.~{R{\"o}hser}, J.~{Kerp}, N.~{Ben Bekhti} and B.~{Winkel},
  \emph{{High-resolution HI and CO observations of high-latitude
  intermediate-velocity clouds}},
  \href{https://doi.org/10.1051/0004-6361/201526801}{\emph{Astronomy \&
  Astrophysics} {\bfseries 592} (2016) A142}
  [\href{https://arxiv.org/abs/1607.00912}{{\ttfamily 1607.00912}}].

\bibitem{Haffner_2003}
L.M.~Haffner, R.J.~Reynolds, S.L.~Tufte, G.J.~Madsen, K.P.~Jaehnig and
  J.W.~Percival, \emph{{The Wisconsin H-alpha Mapper Northern Sky Survey}},
  \href{https://doi.org/10.1086/378850}{\emph{Astrophys. J. Suppl.} {\bfseries
  149} (2003) 405}.

\bibitem{PDG:2022}
{\scshape Particle Data Group} collaboration, \emph{{Review of Particle
  Physics}}, \href{https://doi.org/10.1093/ptep/ptac097}{\emph{PTEP} {\bfseries
  2022} (2022) 083C01}.

\bibitem{2015A&A...578A..87S}
O.~{Snaith}, M.~{Haywood}, P.~{Di Matteo}, M.D.~{Lehnert}, F.~{Combes},
  D.~{Katz} et~al., \emph{{Reconstructing the star formation history of the
  Milky Way disc(s) from chemical abundances}},
  \href{https://doi.org/10.1051/0004-6361/201424281}{\emph{Astronomy \&
  Astrophysics} {\bfseries 578} (2015) A87}
  [\href{https://arxiv.org/abs/1410.3829}{{\ttfamily 1410.3829}}].

\bibitem{Numpy:2020}
C.R.~Harris et~al., \emph{Array programming with {NumPy}},
  \href{https://doi.org/10.1038/s41586-020-2649-2}{\emph{Nature} {\bfseries
  585} (2020) 357}.

\bibitem{Scipy:2020}
P.~Virtanen et~al., \emph{{{SciPy} 1.0: Fundamental Algorithms for Scientific
  Computing in Python}},
  \href{https://doi.org/10.1038/s41592-019-0686-2}{\emph{Nature Methods}
  {\bfseries 17} (2020) 261}.

\bibitem{Karwin:2019}
C.M.~Karwin, S.~Murgia, S.~Campbell and I.V.~Moskalenko,
  \emph{{\textit{Fermi}-LAT Observations of $\gamma$-Ray Emission toward the
  Outer Halo of M31}},
  \href{https://doi.org/10.3847/1538-4357/ab2880}{\emph{The Astrophysical
  Journal} {\bfseries 880} (2019) 95}
  [\href{https://arxiv.org/abs/1903.10533}{{\ttfamily 1903.10533}}].

\bibitem{Roy:2022}
M.~{Roy} and B.B.~{Nath}, \emph{{Gamma-rays from the circumgalactic medium of
  M31}}, \href{https://doi.org/10.1093/mnras/stac1465}{\emph{Mon. Not. Roy.
  Astron. Soc.} (2022) } [\href{https://arxiv.org/abs/2205.12291}{{\ttfamily
  2205.12291}}].

\bibitem{TibetASgamma}
A.~{Neronov}, D.~{Semikoz} and I.~{Vovk}, \emph{{New limit on high Galactic
  latitude PeV $\gamma$-ray flux from Tibet AS$\gamma$ data}},
  \href{https://doi.org/10.1051/0004-6361/202141800}{\emph{Astronomy \&
  Astrophysics} {\bfseries 653} (2021) L4}
  [\href{https://arxiv.org/abs/2107.06541}{{\ttfamily 2107.06541}}].

\bibitem{IceCube:muon}
{\scshape IceCube} collaboration, \emph{{Improved Characterization of the
  Astrophysical Muon-neutrino Flux with 9.5 Years of IceCube Data}},
  \href{https://doi.org/10.3847/1538-4357/ac4d29}{\emph{Astrophys. J.}
  {\bfseries 928} (2022) 50}
  [\href{https://arxiv.org/abs/2111.10299}{{\ttfamily 2111.10299}}].

\bibitem{Dembinski:2017zsh}
H.P.~Dembinski, R.~Engel, A.~Fedynitch, T.~Gaisser, F.~Riehn and T.~Stanev,
  \emph{{Data-driven model of the cosmic-ray flux and mass composition from 10
  GeV to $10^{11}$ GeV}}, \href{https://doi.org/10.22323/1.301.0533}{\emph{PoS}
  {\bfseries ICRC2017} (2018) 533}
  [\href{https://arxiv.org/abs/1711.11432}{{\ttfamily 1711.11432}}].

\bibitem{IceCube:2019hmk}
{\scshape IceCube} collaboration, \emph{{Cosmic ray spectrum and composition
  from PeV to EeV using 3 years of data from IceTop and IceCube}},
  \href{https://doi.org/10.1103/PhysRevD.100.082002}{\emph{Phys. Rev. D}
  {\bfseries 100} (2019) 082002}
  [\href{https://arxiv.org/abs/1906.04317}{{\ttfamily 1906.04317}}].

\bibitem{2020SCPMA..6309801W}
W.~{Wang}, J.~{Han}, M.~{Cautun}, Z.~{Li} and M.N.~{Ishigaki}, \emph{{The mass
  of our Milky Way}},
  \href{https://doi.org/10.1007/s11433-019-1541-6}{\emph{Science China Physics,
  Mechanics, and Astronomy} {\bfseries 63} (2020) 109801}
  [\href{https://arxiv.org/abs/1912.02599}{{\ttfamily 1912.02599}}].

\bibitem{FB:2012a}
F.~{Guo} and W.G.~{Mathews}, \emph{{The Fermi Bubbles. I. Possible Evidence for
  Recent AGN Jet Activity in the Galaxy}},
  \href{https://doi.org/10.1088/0004-637X/756/2/181}{\emph{Astrophys. J.}
  {\bfseries 756} (2012) 181}
  [\href{https://arxiv.org/abs/1103.0055}{{\ttfamily 1103.0055}}].

\bibitem{FB:2012b}
H.Y.K.~{Yang}, M.~{Ruszkowski}, P.M.~{Ricker}, E.~{Zweibel} and D.~{Lee},
  \emph{{The Fermi Bubbles: Supersonic Active Galactic Nucleus Jets with
  Anisotropic Cosmic-Ray Diffusion}},
  \href{https://doi.org/10.1088/0004-637X/761/2/185}{\emph{Astrophys. J.}
  {\bfseries 761} (2012) 185}
  [\href{https://arxiv.org/abs/1207.4185}{{\ttfamily 1207.4185}}].

\bibitem{FB:2014}
M.V.~{Barkov} and V.~{Bosch-Ramon}, \emph{{Formation of large-scale magnetic
  structures associated with the Fermi bubbles}},
  \href{https://doi.org/10.1051/0004-6361/201322743}{\emph{Astronomy \&
  Astrophysics} {\bfseries 565} (2014) A65}
  [\href{https://arxiv.org/abs/1311.6722}{{\ttfamily 1311.6722}}].

\bibitem{FB:2016}
M.J.~{Miller} and J.N.~{Bregman}, \emph{{The Interaction of the Fermi Bubbles
  with the Milky Way's Hot Gas Halo}},
  \href{https://doi.org/10.3847/0004-637X/829/1/9}{\emph{Astrophys. J.}
  {\bfseries 829} (2016) 9} [\href{https://arxiv.org/abs/1607.04906}{{\ttfamily
  1607.04906}}].

\bibitem{2015MNRAS.450.1604L}
Z.~{Lu} et~al., \emph{{Star formation and stellar mass assembly in dark matter
  haloes: from giants to dwarfs}},
  \href{https://doi.org/10.1093/mnras/stv667}{\emph{MNRAS} {\bfseries 450}
  (2015) 1604} [\href{https://arxiv.org/abs/1406.5068}{{\ttfamily 1406.5068}}].

\bibitem{2015A&A...583A..61G}
B.R.~{Granett} et~al., \emph{{The VIMOS Public Extragalactic Redshift Survey.
  Reconstruction of the redshift-space galaxy density field}},
  \href{https://doi.org/10.1051/0004-6361/201526330}{\emph{Astronomy \&
  Astrophysics} {\bfseries 583} (2015) A61}
  [\href{https://arxiv.org/abs/1505.06337}{{\ttfamily 1505.06337}}].

\bibitem{2016:GalNu}
M.~{Ahlers}, Y.~{Bai}, V.~{Barger} and R.~{Lu}, \emph{{Galactic neutrinos in
  the TeV to PeV range}},
  \href{https://doi.org/10.1103/PhysRevD.93.013009}{\emph{Phys. Rev. D}
  {\bfseries 93} (2016) 013009}
  [\href{https://arxiv.org/abs/1505.03156}{{\ttfamily 1505.03156}}].

\bibitem{2020MNRAS.498.3125P}
R.~{Pakmor}, F.~{van de Voort}, R.~{Bieri}, F.A.~{Gomez}, R.J.J.~{Grand},
  T.~{Guillet} et~al., \emph{{Magnetizing the circumgalactic medium of disc
  galaxies}}, \href{https://doi.org/10.1093/mnras/staa2530}{\emph{MNRAS}
  {\bfseries 498} (2020) 3125}
  [\href{https://arxiv.org/abs/1911.11163}{{\ttfamily 1911.11163}}].

\end{thebibliography}\endgroup

\end{document}